\documentclass[final]{aipproc}

\layoutstyle{8x11double}
\usepackage{amsmath}

\newcommand{\gluino}{\tilde{g}}
\newcommand{\squark}{\tilde{q}}
\newcommand{\stopp}{\tilde{t}}

\begin{document}

\title{Hadronic Production of Colored SUSY Particles \\ with Electroweak NLO Contributions}

\classification{12.15.Lk, 14.80Ly}
\keywords{Supersymmetric Standard Model, NLO Computations, Hadronic Colliders}

\author{Wolfgang Hollik}{
  address={Max-Planck-Institut f\"ur Physik, 
 F\"ohringer Ring 6, D-80805 M\"unchen, Germany}
}

\author{Monika Kollar}{
  address={Max-Planck-Institut f\"ur Physik, 
 F\"ohringer Ring 6, D-80805 M\"unchen, Germany}
}

\author{Edoardo Mirabella}{
  address={Max-Planck-Institut f\"ur Physik, 
 F\"ohringer Ring 6, D-80805 M\"unchen, Germany}
}

\author{Maike K. Trenkel}{
  address={Max-Planck-Institut f\"ur Physik, 
 F\"ohringer Ring 6, D-80805 M\"unchen, Germany}
}

\begin{abstract}
We consider the production of squarks and gluinos at hadronic colliders. An overview over 
the class of processes is given. 
We investigate in detail the tree-level and higher order EW contributions to the cross sections. Special care has to be taken to obtain infrared finite observables. We study numerically stop--anti-stop and squark--gluino production at the LHC.
\end{abstract}

\maketitle


\section{Introduction}

If SUSY is realized, 
colored SUSY particles will provide large production 
cross sections at hadron colliders since they are produced via the strong 
interaction. Pair production of squarks and gluinos is, therefore, among the most promising 
SUSY discovery channels. 

The cross sections are substantially altered by higher order corrections. 
NLO QCD corrections at  $\mathcal{O}(\alpha_s^3)$ range typically at 20-30\% \cite{Beenakker:1996ch} or even larger values for top-squark pair production 
 \cite{Beenakker:1997ut}. It was shown that they reduce the 
factorization and renormalization scale dependence considerably.
Only recently also NLO EW corrections of $\mathcal{O}(\alpha_s^2 \alpha)$
 were considered, to top-squark pair production~\cite{Hollik:2007wf,Beccaria:2008mi}, 
 to pair production of squarks of the first generations~\cite{Hollik:2008}, and
 to squark--gluino production~\cite{Hollik:2008sqg}. 

The aim of this talk is to provide an overview over the class of processes and to 
illustrate the effects of the EW contributions. 
The focus will be on top-squarks (stops~$\stopp$), candidates for the lightest squark 
within many SUSY models owing to the large top-Yukawa coupling~\cite{Ellis:1983ed}, 
and on squark--gluino ($\squark\gluino$) final states, constituting an important fraction of 
colored SUSY particle production processes in a wide range of the parameter space.
We investigate in detail the various EW contributions to the cross sections,
including tree-level $\mathcal{O}(\alpha^2 + \alpha_s \alpha)$ and NLO $\mathcal{O}(\alpha_s^2 \alpha)$ corrections. 
Special care has to be taken to obtain infrared (IR) finite observables. 
We present numerical results for $\stopp\stopp^*$ and $\squark\gluino$~production
within the specific SUSY scenario SPS1a$'$~\cite{AguilarSaavedra:2005pw}.


\section{Tree-level contributions} 

At hadron colliders, pair production of squarks and gluinos 
proceeds at lowest oder QCD via the following partonic processes.

-- Gluino pairs are produced by $gg$ and $q\bar{q}$~initial states.

-- Squark--gluino final states require quark--gluon initial states, the quark and the produced 
squark being of the same flavor. Accordingly, $\tilde{t}\,\gluino$~final states cannot be produced at LO
owing to the vanishing top-quark PDF, and $\tilde{b}\,\gluino$~production is suppressed. Furthermore, bottom-squarks are experimentally distinguishable from other squarks
by their decay products. We therefore restrict the discussion of $\squark\gluino$~production to left- and right-handed (anti-)squarks of the first two generations.

-- Squark--anti-squark pairs are produced via $gg$~fusion and $q\bar{q}$~annihilation. 
The $q\bar{q}$~initiated processes can be either gluon-mediated $s$-channel diagrams (with quarks of any flavor in the initial state) or gluino-mediated $t$-channel diagrams. The latter also allow for $\squark\squark$~production, which will be discussed elsewhere.

-- Stop--anti-stop (and sbottom--anti-sbottom) pair production has to be discussed separately from  $\squark\squark^*$~production since, excluded by the PDF, it does not proceed via $t$-channel diagrams. Moreover, squarks of the third generation are experimentally distinguishable and 
the L--R-mixing has to be taken into account. Due to the absence of the $t$-channel diagram, stops (and sbottoms) can only be produced diagonally at LO.
\medskip

These production mechanisms of $\mathcal{O}(\alpha_s^2)$ are the dominant ones for 
squark and gluino production at hadronic colliders. 
Besides, diagonal and non-diagonal squark pairs can also be produced by 
$q\bar{q}$~induced tree-level EW processes \cite{Bornhauser:2007bf,Bozzi:2007me}.
For stops and sbottoms, only $s$-channel diagrams with $\gamma$ or $Z$~exchange are 
present at $\mathcal{O}(\alpha^2)$  and give negligible contributions.
However for squarks of the first generations also $t$-channel diagrams mediated by
 $\tilde{\chi}^0$ or $\tilde{\chi}^{\pm}$~exchange contribute and, additionally, 
non-zero interferences of 
$\mathcal{O}(\alpha_s \alpha)$ between the QCD and EW diagrams  arise.
 
Finally, photon-induced processes are further tree-level channels that contribute 
to (top-)squark pair production (via photon--gluon fusion) and to squark--gluino 
production (via photon--quark fusion). Although these channels are in general
suppressed by the photon distribution in the proton which is only 
non-zero at NLO QED, they can become sizable as pointed out in
 \cite{Hollik:2007wf,Hollik:2008}.


\section{EW NLO corrections}

One subset of EW NLO corrections are the virtual corrections. 
They arise from the interference between tree-level QCD diagrams and diagrams 
with one-loop EW insertions. 
In case of $q\bar{q}$~initial states, further $\mathcal{O}(\alpha_s^2 \alpha)$~corrections 
arise from the interference of $\mathcal{O}(\alpha)$~EW
tree-level and $\mathcal{O}(\alpha_s^2)$~QCD box diagrams.

Getting an UV finite result requires renormalization of the involved quarks
and squarks, which has been performed in the on-shell scheme.
In case of $\stopp\stopp^*$~production, it is not necessary 
to renormalize the gluon field and the strong coupling $\alpha_s$ at this order of 
perturbation theory. Similarly for $\squark\gluino$~production, also the gluino field needs 
not to be renormalized. The situation is more involved for $\squark\squark^*$~production, 
where the full one-loop QCD amplitude enters through interference contributions 
with the (enlarged) EW tree-level diagrams~\cite{Hollik:2008}.

As a second type of singularities, one has to deal with IR (soft) singularities that arise
from loop diagrams involving virtual photons. 
An IR finite result is obtained only in the sum of virtual corrections and 
real photon radiation processes. The diagrams are regularized by 
introducing a fictitious photon mass. 
For $\squark\squark^*$ and $\stopp\stopp^*$~production, also 
virtual gluons arise and become singular in the soft limit. To obtain an IR finite result, 
real gluon radiation processes have to be included at the appropriate order.
The gluonic IR singularities are Abelian-like and 
can similarly be regularized by a finite gluon mass.

In case of a light quark in the initial state, also collinear singularities 
occur if the quark radiates off a photon in the collinear limit.
We therefore keep the initial state quark masses as regulators where necessary.
Parts of the singularities drop out in the sum of virtual and real corrections. 
Single logarithms owing to collinear singularities, however, survive and 
have to be absorbed into the quark parton distribution functions (PDFs). 
This can formally be achieved by a redefinition of the PDFs at NLO QED, 
in complete analogy to factorization for NLO QCD 
 calculations~\cite{Baur:1998kt,Diener:2003ss}.

As mentioned before, $\stopp\stopp^*$~production at 
$\mathcal{O}(\alpha_s^2 \alpha)$ also includes interference contributions 
from tree-level EW diagrams and one-loop QCD boxes. Providing an additional 
source of soft gluon singularities, these terms are essential in the sum 
of virtual and real gluon corrections in order to obtain an IR finite result.
In case of $\squark\squark^*$~production the real gluon radiation also
exhibits collinear singularities and the redefinition of the PDFs has 
to be done accordingly.

Finally, real (anti-)quark radiation processes
give contributions of $\mathcal{O}(\alpha_s^2 \alpha)$ 
through the interference of EW mediated and QCD mediated 
diagrams. For $\stopp\stopp^*$ and $\squark\gluino$~production, 
they are IR and collinear finite. In some diagrams, internal-state SUSY particles 
can go on-shell and the propagators are regularized by inserting the 
width of the respective particle. 
For $\stopp\stopp^*$~production, however, 
only SM particles appear as internal particles and real quark radiation 
is negligible and, therefore, not included in the numerical analysis.

\section{Numerical Results}

For the numerical discussion we refer to the mSUGRA scenario SPS1a$'$~\cite{AguilarSaavedra:2005pw}
and use {\tt Softsusy}~\cite{Allanach:2001kg} to evolve the GUT 
scale parameters down to lower scales. Factorization 
and renormalization scale are set equal to 1~TeV. A translation of the 
$\overline{\text{DR}}$~parameters into the on-shell scheme 
is achieved by relating the renormalized masses at the one-loop 
level, leading to $m(\stopp_1) = 359.6$~GeV as the mass of the lighter stop.
The top-quark mass is set to $170.9~$GeV, all other 
Standard Model parameters are chosen in accordance with 
\cite{AguilarSaavedra:2005pw}.
As discussed above, we need a set of PDFs that includes NLO QED corrections,
as provided by MRST\,2004\,QED~\cite{Martin:2004dh}.

\begin{table}
\begin{tabular}{ccccc}
\hline
\tablehead{1}{c}{c}{final state}	
  & \tablehead{1}{c}{c}{$\sigma^{\rm LO}$}
  & \tablehead{1}{c}{c}{$\sigma^{\rm EW}_{\alpha_s^2 \alpha}$}
  & \tablehead{1}{c}{c}{$\sigma^{\rm EW}_{\alpha_s \alpha}$}
  & \tablehead{1}{c}{c}{$\delta^{\rm EW}$}   \\
\hline
$\stopp_1 \stopp_1^*$  & 1830~fb & $-15.0$~fb & 34.1~fb   & 1.0\% 
\\[.5ex]
\hline
$\tilde{u}_R\gluino + \tilde{d}_R\gluino$ & 8900~fb & 14.6~fb & 5.05~fb  & 0.22\% 
\\[.5ex]
$\tilde{u}_L\gluino + \tilde{d}_L\gluino$ & 8220~fb & $-197$~fb & 4.62~fb  & $-2.3\%$
\\[.5ex]
$\squark\gluino$ & 17120~fb & $-183$~fb & 9.67~fb  & $-1.0\%$\\
\hline
\end{tabular}
\caption{Integrated cross sections for $\stopp_1 \stopp_1^*$ and 
$\squark\gluino$ production at the LHC within the SPS1a$'$ scenario.}
\label{tab_numbers}
\end{table}

In Table~\ref{tab_numbers} we present the integrated cross 
sections for $\stopp_1 \stopp_1^*$ and $\squark\gluino$~production at the LHC. 
Given are the LO, the $\mathcal{O}(\alpha_s^2\alpha)$,
and the photon-induced $\mathcal{O}(\alpha_s \alpha)$~cross sections,
and the complete EW contribution relative to the LO result, 
in obvious notation.
For $\stopp_1 \stopp_1^*$~production, the photon-induced processes 
dominate over the NLO $\mathcal{O}(\alpha_s^2\alpha)$~corrections 
and change the result substantially. 
The complete EW contribution is small and positive~(1\%).
For $\squark\gluino$~final states, we present cross sections
 for left-handed and right-handed squarks separately and also give 
the inclusive results. Squarks of the second generation and 
charge conjugated processes are summed over, 
differing only by the required PDF. 
The photon-induced channels are less important than for stops,
mainly suppressed by the parton density of the initial quark
which is rather sensitive to the high final state masses 
[\mbox{$m(\tilde{u}_R) = 543.4$~GeV,}
\mbox{$m(\tilde{u}_L) = 560.7$~GeV,} \mbox{$m(\tilde{d}_R) = 539.4$~GeV,} 
$m(\tilde{d}_L)  = 566.4$~GeV, $m(\tilde{g})=  609.0$~GeV].
Note that $\sigma^{\rm EW}_{\alpha_s \alpha}$ is independent on the 
chirality of the produced squark. Virtual and real corrections, of 
course, are not and $\sigma^{\rm EW}_{\alpha_s^2 \alpha}$ 
is only important for left-handed squarks. 
The total EW contribution amounts~$-1\%$, for inclusive $\squark\gluino$ production.

\begin{figure}
  \includegraphics[width=\columnwidth]{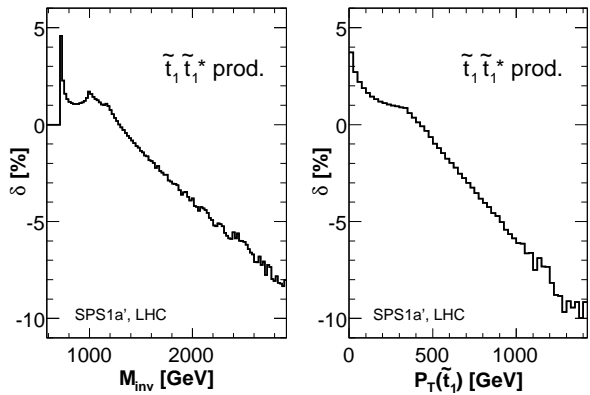}
  \caption{EW contribution relative to the LO cross section for 
 $\stopp_1\stopp^*_1$ production at the LHC. Shown are the distributions with 
respect to the invariant mass of the $\stopp_1\stopp^*_1$ pair and to 
the transverse momentum of one of the stops.}
 \label{fig_stop}
\end{figure}

\begin{figure}
  \includegraphics[width=\columnwidth]{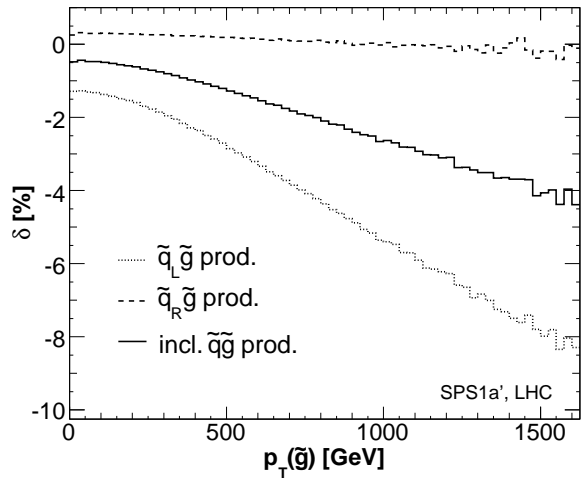}
  \caption{EW contribution relative to the LO cross section for 
 $\squark\gluino$ production at the LHC as distribution with 
respect to the transverse momentum of the gluino.}
  \label{fig_squarkgluino}
\end{figure}

We illustrate the numerical impact of the EW contribution 
on the LO cross section in Fig.~\ref{fig_stop} and 
Fig.~\ref{fig_squarkgluino}.

Fig.~\ref{fig_stop} refers to $\stopp_1\stopp^*_1$~production and 
shows the relative EW contribution as distribution 
with respect to the invariant mass ($M_{\rm inv}$) 
of the $\stopp_1\stopp^*_1$~pair and 
to the transverse momentum ($p_T$) of one of the stops.
In the $M_{\rm inv}$~distribution, threshold effects arise 
from stop and sbottom pairs in the loop diagrams.
In total, the EW contributions grow in size for increasing 
$M_{\rm inv}$ and $p_T$ and reach about $-10\%$ for 
$M_{\rm inv}\sim 3$~TeV and $p_T\sim1.5$~TeV.

Fig.~\ref{fig_squarkgluino} refers to $\squark\gluino$~production.
Again, we consider left- and right-handed squarks separately and give 
also the inclusive results. Presented is the relative EW contribution 
with respect to $p_T$ of the produced gluino.
Differences for $p_T(\squark)$~are expected to be small, 
only arising from real photon and real quark radiation processes.
For left-handed squarks, the EW contribution
 reaches the $10\%$~level in the high-$p_T$ range.
It is negligible, however, for right-handed squarks and 
the distribution is almost flat. 
 As a consequence, the EW contribution to inclusive 
$\squark\gluino$~production is moderate, 
about $-5\%$ for $p_T\sim1.5$~TeV.


\section{Conclusions}

We considered the hadronic production of colored SUSY particles, 
squarks and gluinos, and classified the various processes.
We focused on the EW contribution, 
due to tree-level $\mathcal{O}(\alpha_s  \alpha)$~processes 
and NLO $\mathcal{O}(\alpha_s^2 \alpha)$~corrections.
We numerically investigated $\stopp_1\stopp^*_1$ and 
$\squark\gluino$~production at the LHC within the SPS1a$'$~scenario.
The EW contribution to the total cross sections is small, however it 
reaches (in size) the $10-20\%$~level in the 
$p_T$ and invariant mass distributions and is thus significant.


\bibliographystyle{aipproc}  
\bibliography{references}

\end{document}